\documentclass{article}
\begin{document}
\title{On the possible new heavy scalar and pseudoscalar resonances at the LHC}
\author{Davor Palle \\
ul. Ljudevita Gaja 35, 10000 Zagreb, Croatia \\
email: davor.palle@gmail.com}
\maketitle

\begin{abstract}
We argue that the possible new heavy boson resonance of 750 GeV
is an ideal candidate as a twin particle of the 125 GeV scalar
boson, both emerging from the large mixing of the scalar toponium and
scalar gluonium.
The twin pseudoscalar particles are expected to have smaller masses.
The discovery of the 750 GeV
resonance is possible only with a much more data than for the 125 GeV resonance since only
the gluonium component is detectable above the toponium threshold.
The similar type of the QCD resonances is expected in the bottomonium-gluonium
system. If the LHC will discover all these heavy quarkonium-gluonium resonances,
the absence of the Higgs scalar should not be considered an obstacle because
the nonsingular theory with the UV cutoff fixed
by the weak boson masses is superior to the Standard Model. Namely, it solves the basic
problems for the SM such as:
(1) the existence of light neutrinos, (2) dark matter particles to be
the heavy Majorana neutrinos and (3) broken lepton and
baryon numbers. 
\end{abstract}
\vspace{6mm}

We are witnessing the great discovery potential of the Large
Hadron Collider (LHC). The Run 1 experiments of the LHC 
at 7 and 8 TeV center of mass energy found new exotic hadrons interpreted as
tetraquark \cite{LHC1} or  pentaquark \cite{LHC2} states.

The special attention should be devoted to the discovery of the
125 GeV boson resonance \cite{LHC3}. It is established that it is
indeed a scalar particle. Owing to the fact that the SM Higgs scalar cannot 
generate neutrino masses, one has to expand the scalar sector of the model
if we adopt the approach that the Higgs mechanism is responsible for the generation of
masses.
On the other hand, the overall fit of the electroweak data of
the LEP1, LEP2, SLC, etc. with the SM radiative corrections results in the mass of the 
Higgs $m_{H} =  89^{+22}_{-18} GeV$ \cite{PDG}. 
The nonperturbative stability analysis of the SM Higgs sector requires much heavier
Higgs mass $m_{H} >  180 GeV$ \cite{Talijani}.
The measurements of the partial decay widths of the scalar 125 GeV resonance
are too far from the scientific golden standard of precision to be considered
compatible with the SM Higgs couplings. 

Despite all these facts, the 125 GeV resonance is proclaimed to be the SM Higgs
particle.
Recently, the ATLAS and the CMS collaborations announced the possible discovery 
of the new 750 GeV heavy boson decaying into two photons \cite{LHC4}.

Immediately after the discovery of the 125 GeV resonance, P. Cea
suggested that this resonance could be the QCD bound state
as a mixture of toponium and gluonium \cite{Cea}.
However, this interpretation implies the existence of two heavy bosons. Let us write
the corresponding mass matrix \cite{Cea,Glashow}:

\begin{eqnarray*}
M=\left( \begin{array}{cc}
m_{gg} + A  & A \\
A  & m_{t\bar{t}} + A \end{array}
\right).
\end{eqnarray*}

Note that this mass matrix is an exact mass matrix for any heavy quarkonium-gluonium
system in the theory of the Bethe-Salpeter equations \cite{Nakanishi}.
The eigenvalue problem is reduced to the following algebraic system of the three
nonlinear equations:

\begin{eqnarray*}
\sin\theta\cos\theta (m_{gg}-m_{t\bar{t}})+(\cos^{2}\theta-\sin^{2}\theta)A&=&0, \\
\cos^{2}\theta\ m_{gg}+\sin^{2}\theta\ m_{t\bar{t}}+A (1-2\sin\theta\cos\theta)&=&m_{1}, \\
\cos^{2}\theta\ m_{t\bar{t}}+\sin^{2}\theta\ m_{gg}+A (1+2\sin\theta\cos\theta)&=&m_{2},
\end{eqnarray*}
\begin{eqnarray*}
m_{gg}=gluonium\ mass,\ m_{t\bar{t}}=toponium\ mass,\ m_{1}=lighter\ twin\ mass, \\
m_{2}=heavier\ twin\ mass,\ \theta=mixing\ angle,\ A=annihilation\ matrix\ element,
\end{eqnarray*}
\begin{eqnarray*}
|1>=\cos\theta\ |gg>-\sin\theta |t\bar{t}>,\ |2>=\sin\theta\ |gg>+\cos\theta |t\bar{t}>.
\end{eqnarray*}

There are six variables in the system - therefore we can fix three variables and solve 
the system to find the remaining three. 

Besides some hints for the 750 GeV resonance, we are witnessing to some evidence
for the 96 GeV resonance \cite{Barate} and the 28 GeV resonance \cite{Heister}.
Solutions of the above algebraic system for heavy quarkonium-gluonium system give us
the following twin particles:

\begin{eqnarray*}
&&SCALARS: \\
&&m_{gg}=1.65 GeV,\ m_{t\bar{t}}=364 GeV,\ m_{1}=125 GeV \\
&&\Rightarrow m_{2}=750.5 GeV,\ A=254.9 GeV,\ \sin\theta=0.459,\ \frac{m_{2}}{A}=2.94 \\
&& --------------------------- \\
&&m_{gg}=1.65 GeV,\ m_{b\bar{b}}=9.86 GeV,\ m_{1}=5 GeV \\
&&\Rightarrow m_{2}=28.07 GeV,\ A=10.78 GeV,\ \sin\theta=0.568,\ \frac{m_{2}}{A}=2.60 \\
&& --------------------------- \\
&&PSEUDOSCALARS: \\
&&m_{gg}=2.5 GeV,\ m_{t\bar{t}}=321 GeV,\ m_{1}=96 GeV \\
&&\Rightarrow m_{2}=547.5 GeV,\ A=160.0 GeV,\ \sin\theta=0.384,\ \frac{m_{2}}{A}=3.42\ .
\end{eqnarray*}

The annihilation term A is large owing to the multigluon
strong coupled exchange in the quantum loop. The saturation should be
expected for strong interactions on the high top quark (bottom quark)-gluon ladder.
The vector and tensor heavy quarkonium-gluonium systems can not be excluded. Note that
the Landau-Yang theorem in QCD is not valid, thus the vector particle can couple to two gluons.

It is well known that the toponium states decay quickly via weak interactions and therefore cannot
be observed by the LHC detectors. This is the reason why heavier twin meson has yet
to be observed, and it will prove to be difficult because it consists mainly of toponium.

There is a serious theoretical challenge ahead to evaluate the annihilation matrix elements in
both scalar and pseudoscalar channels by solving Bethe-Salpeter equations or within the QCD on the
lattice. However, the QCD lattice calculations with b-quarks are feasible. 

The possibility that we could be left without the Higgs scalar should not pose
as a matter of concern. Namely, the Higgs mechanism built into the electroweak theory helped 
to establish the SM model, but does not solve the problem of masses of the 
elementary particles, i.e. Higgs potential and Yukawa couplings are free parameters.
However, we know that the lepton and quark masses fulfil profound patterns: only three
fermion families; characteristic mass gaps; quarks heavier than leptons within the same
family, and very light neutrinos.

The resolution of these problems requires the introduction of a new paradigm.
The theory of noncontractible space and its consequence on the relations between
gauge, conformal and discrete symmetries are explained in ref. \cite{Palle1}.
The masses of elementary particles are mass singularities of propagator Green 
functions which are solutions to the nonsingular Dyson-Schwinger equations.
The theory contains three light and three heavy Majorana neutrinos \cite{Palle2}. The lepton
and baryon numbers are broken \cite{Palle1,Japovi}.
The impact of the theory on the phenomenology of the rare B-meson processes and
the anomalous magnetic moment of the muon 
can be found in
ref. \cite{Palle3}, whereas the effect on the strong interactions - strong coupling, spin asymmetry in the
single t-quark production or t-quark charge asymmetry, etc. in ref. \cite{Palle4}.

The probable violation of the lepton universality in the semileptonic decays is probably a result of
the uncertain and questionable evaluations of the hadron matrix elements with heavy b-quark, and not
the signal of new physics.
The pure leptonic decays of the B mesons could be
more interesting \cite{Palle3} since one needs to evaluate the matrix element
with only one meson state. Note that the similar problems appear in
the LEP observables \cite{Baak} and in the radiative corrections of the W boson mass
\cite{CDF,Palle9}. The role of the massless unphysical $\zeta$ particle introduced instead
of the Higgs boson is essential in the longitudinally polarized W boson scattering
\cite{New1}.

The essential relation for the cancellation of the global $SU(2)$ anomaly
\cite{Palle1} valid for the Weinberg angle and Dirac fermion mixing angles:

\begin{eqnarray*}
\Theta_{W}=2 (\Theta_{12}^{D}+\Theta_{23}^{D}+\Theta_{31}^{D}),
\end{eqnarray*}

must be valid even for Majorana neutrinos. However, in the case of the inverted
mass hierarchy, for example if $m_{\nu,1}^{M} > m_{\nu,2}^{M}$ for 
light Majorana neutrinos,
$m_{N,1}^{M} < m_{N,2}^{M}$ for heavy Majorana neutrinos
and $m_{\nu,1}^{D} < m_{\nu,2}^{D}$ for Dirac neutrinos, the see-saw mechanism and the Euler
matrix imply $\Theta_{12}^{D}=-\Theta_{12}^{M}$:

\begin{eqnarray*}
\left( \begin{array}{cc}
\cos \Theta_{12} & \sin \Theta_{12} \\
-\sin \Theta_{12}  & \cos \Theta_{12} \end{array}
\right)
\left( \begin{array}{c}
u_{1} \\
u_{2} \end{array}
\right)
\Leftrightarrow
\left( \begin{array}{cc}
\cos \Theta_{12} & -\sin \Theta_{12} \\
\sin \Theta_{12}  & \cos \Theta_{12} \end{array}
\right)
\left( \begin{array}{c}
u_{2} \\
u_{1} \end{array}
\right).
\end{eqnarray*}

The present knowledge
of the neutrino mixing matrix \cite{PDG1} and the above cancellation
condition favour the inverted mass hierarchy.

The H.E.S.S. source J1745-290 discovered in 2004 \cite{Aharonian} is a prefect candidate 
as a source of the very heavy cold dark matter particle (possibly heavy Majorana
neutrinos) \cite{Cembranos}.

The connection and the universality of the theory of noncontractible space with the Einstein-Cartan
cosmology can be examined in ref. \cite{Palle5}. The heavy Majorana neutrinos are candidates
for cold dark matter particles and the angular momentum of the Universe is the dark energy
\cite{Palle6}. The right-handed rotation of the Universe is an inevitable consequence of the left-handed
weak interactions \cite{Palle7}.
The consequences of the Einstein-Cartan cosmology on the high-redshift Universe can be found in
the ref. \cite{Palle8}.

We emphasize two new important problems resolved by the Einstein-Cartan cosmology:
(1) the $S_{8}$ problem \cite{New2} and (2) the matter kinematic dipole problem \cite{New3}.


\begin{thebibliography}{100}

\bibitem{LHC1} LHCb Collab., Phys. Rev. Lett. {\bf 112}, 222002 (2014) [arXiv:1404.1903].

\bibitem{LHC2} LHCb Collab., Phys. Rev. Lett. {\bf 115}, 072001 (2015) [arXiv:1507.03414].

\bibitem{LHC3} ATLAS Collab., Phys. Lett. {\bf B 716}, 1 (2012) [arXiv:1207.7214];
               CMS Collab., Phys. Lett. {\bf B 716}, 30 (2012) [arXiv:1207.7235].

\bibitem{PDG} J. Erler and A. Freitas (Particle Data Group), Chin. Phys. {\bf C 38}, 090001 (2014).

\bibitem{Talijani} G. Degrassi et al., JHEP {\bf 08}, 098 (2012) [arXiv:1205.6497].

\bibitem{LHC4} M. Kado (ATLAS Collab.), "Results with the Full 2015 Data Sample from the ATLAS
               experiment", presented at CERN, December 15, 2015;
               J. Olsen (CMS Collab.), "CMS 13 TeV Results", presented at CERN, December 15, 2015.

\bibitem{Cea} P. Cea, "Comment on the evidence of the Higgs boson at LHC", arXiv:1209.3106.

\bibitem{Glashow} \'{A}. de R\'{u}jula, H. Georgi and S. L. Glashow, 
                  Phys. Rev. {\bf D 12}, 147 (1975).

\bibitem{Nakanishi} N. Nakanishi, Prog. Theor. Phys. Suppl. {\bf 43}, 1 (1969).

\bibitem{Barate} R. Barate et al. (LEP Working Group), Phys. Lett. {\bf B 565}, 61 (2003)
                 [hep-ex:0306033]; A. M. Sirunyan et al. (CMS Collab.)
                 Phys. Lett. {\bf B 793}, 320 (2019) [arXiv:1811.08459].
                 
\bibitem{Heister} A. Heister, "Observation of an eccess at 30 GeV in the opposite
                  sign di-muon spectra of $Z\rightarrow b\bar{b}+X$ events recorded by the
                  ALEPH experiment at LEP", arXiv:1610.06536;                
                  CMS Collab., JHEP {\bf 11}, 161 (2018) [arXiv:1808.01890].

\bibitem{Palle1} D. Palle, Nuovo Cim. {\bf A 109}, 1535 (1996) [hep-ph/9706266].

\bibitem{Palle2} D. Palle, Nuovo Cim. {\bf B 115}, 445 (2000) [hep-ph/9910512];
                 D. Palle, "On Dyson-Schwinger equations and the number of fermion
                 families", arXiv:hep-ph/0703203.

\bibitem{Japovi} M. Fukugita and T. Yanagida, Phys. Lett. {\bf B 174}, 45 (1986).

\bibitem{Palle3} D. Palle, Acta Phys. Pol. {\bf B 43}, 1723 (2012) [arXiv:1111.1638];
                 D. Palle, "On the rare $B_{s}$ to two muons decay and noncontractibility
                 of the physical space", arXiv:1111.1639;
                 D. Palle, "On the quantum loop suppressed electroweak processes", 
                 arXiv:1210.4404; D. Palle, Acta Phys. Pol. {\bf B 47}, 1237 (2016)
                 [arXiv:1601.07781].
                 
\bibitem{Palle4} D. Palle, Hadronic J. {\bf 24}, 87 (2001) [hep-ph/9804326];
                 D. Palle, Acta Phys. Pol. {\bf B 43}, 2055 (2012) [arXiv:1204.1171].
                 
\bibitem{Baak} M. Baak and R. Kogler, "The global electroweak Standard Model fit after
               the Higgs discovery", arXiv:1306.0571.
               
\bibitem{CDF} CDF Collab., Science {\bf 376}, 170 (2022).

\bibitem{Palle9} D. Palle, "W boson mass anomaly and noncontractibility 
                 of the physical space", arXiv:2302.10234.

\bibitem{New1} D. Palle,
 "Longitudinally polarized same-sign W boson pairs scattering and
   noncontractibility of the physical space",
 in Croatian Scientific Bibliography: CROSBI ID 941869 (2026).

\bibitem{PDG1} Particle Data Group, https://pdg.lbl.gov. 

\bibitem{Aharonian} F. Aharonian et al. (H.E.S.S. Collab.), A$\&$A {\bf 425}, L13 (2004)
                    [astro-ph/0408145].
                    
 \bibitem{Cembranos} J. A. R. Cembranos, V. Gammaldi and A. L. Maroto,
                     JCAP {\bf 04}, 051 (2013) [arXiv:1302.6871].

\bibitem{Palle5} D. Palle, Nuovo Cim. {\bf B 111}, 671 (1996) [astro-ph/9706012];
                 D. Palle, Nuovo Cim. {\bf B 114}, 853 (1999) [astro-ph/9811408];
                 D. Palle, Nuovo Cim. {\bf B 122}, 67 (2007) [astro-ph/0604287].

\bibitem{Palle6} D. Palle, Eur. Phys. J. {\bf C 69}, 581 (2010) [arXiv:0902.1852];
                           ZhETF {\bf 145}, 671 (2014) [arXiv:1405.3435].

\bibitem{Palle7} D. Palle, Entropy {\bf 14}, 958 (2012) [arXiv:0802.2060].

\bibitem{Palle8} D. Palle, "Einstein-Cartan cosmology and the high-redshift Universe",
                 arXiv:2106.08136; D. Palle, "Einstein-Cartan cosmology and the 
                 CMB anisotropies", arXiv:2204.10283.

\bibitem{New2} D. Palle,
 "Einstein-Cartan cosmology and the $S_{8}$ problem",
 arXiv:2502.20425.

\bibitem{New3} D. Palle,
 "Einstein-Cartan cosmology and the matter kinematic dipole anomaly",
 in Croatian Scientific Bibliography: CROSBI ID 942689 (2026).

\end{thebibliography}
\end{document}